\RequirePackage{fix-cm}
\documentclass[smallextended]{svjour3}       % onecolumn (second format)
\smartqed  % flush right qed marks, e.g. at end of proof
\usepackage{graphicx}
\usepackage{epstopdf} 
\usepackage{multirow}
\usepackage{float}
\usepackage[table]{xcolor}
\usepackage[backend=bibtex,sorting=none]{biblatex}

\begin{document}

\title{HMM and DTW for Evaluation of therapeutical gestures using kinect%\thanks{Grants or other notes
}


\author{Carlos Palma         \and
       Augusto Salazar \and
			Francisco Vargas
			%etc.
}


\institute{Carlos Palma  \at
              Grupo SISTEMIC, Facultad de Ingenier\'ias Universidad de Antioquia UdeA \\
							Calle 70 No. 52 - 21, Medell\'in, Colombia.\\
              \email{carlos.palma@udea.edu.co}           %  \\
           \and
           Augusto Salazar\at
              Grupo SISTEMIC, Facultad de Ingenier\'ias Universidad de Antioquia UdeA \\
							Calle 70 No. 52 - 21, Medell\'in, Colombia.\\
							Grupo de investigaci\'on AEyCC, Facultad de Ingenier\'ias, Instituto Tecnol\'ogico Metropolitano ITM\\
							Carrera 21 No. 54-10, Medell\'in, Colombia.\\
							\email{augusto.salazar@udea.edu.co} 
							\and
							Francisco Vargas \at
							  Grupo SISTEMIC, Facultad de Ingenier\'ias Universidad de Antioquia UdeA \\
								Calle 70 No. 52 - 21, Medell\'in, Colombia.\\
							\email{jesus.vargas@udea.edu.co} 
}

\date{Received: date / Accepted: date}

\maketitle

\begin{abstract}
Automatic recognition of the quality of movement in human beings is a challenging task, given the difficulty both in defining the constraints that make a movement correct, and the difficulty in using noisy data to  determine if these constraints were satisfied.
This paper presents a method for the detection of deviations from the correct form in movements from physical therapy routines based on  Hidden Markov Models, which is compared to Dynamic Time Warping. The activities studied include upper an lower limbs movements, the data used  comes from a Kinect sensor. Correct repetitions of the activities of interest were recorded, as well as deviations from these correct forms. The ability of the proposed approach to detect these deviations was studied. Results show that a system based on HMM is much more likely to determine if a certain movement has deviated from the specification.
\end{abstract}

\section{Introduction}
\label{intro}

Physical therapy is a common step in the rehabilitation process for many injuries and diseases. Movements used in therapy and conditioning routines are well defined in order to strengthen specific body segments and prevent injuries. The usual method by which one of these movements is evaluated is by having a human expert observe it and give feedback to the person executing it. This introduces subjectivity in the evaluation process and implies the need for expert personnel in rehabilitation, increasing the costs for the health systems. Developing a low cost, precise evaluation system for therapeutical gestures could help ease the burden on health systems all over the world while increasing objectivity in the measurements.

 Most of the research on the computer vision community focused on human activities has been directed towards the recognition of activities on video sequences. An extensive review can be found in Mohamed et al\parencite{mohamed2015novice}. In the case of rehabilitation and therapy however the main focus lies on the quality of the movement executed by the patient, and not in the identification of the activity. This means that an automatic system must be able to determine whether the patient has executed the movement according to the specification of an expert.

Several different problems arise however when trying to translate movement specifications into software conditions, as clearly evidenced by the work of Velloso et al \parencite{velloso2013motionma}, where several experts in weightlifting miscalculated the ideal angles to be formed during execution of common exercises.
Even for a clearly defined activity there remains a set of problems to be tackled, like the noise inherent to the measurements taken by the sensors used to monitor the activity, and self-occlusions when using a computer vision system. Another problem lies in the generalization of the evaluation approach to different subjects, given the variability inherent in the human population.

Two widespread techniques used to measure the adjustment of an observation to a pattern of data in a time series are Dynamic Time Warping (DTW) and Hidden Markov Models (HMM). The former consists in aligning a sequence to another of different length by means of finding the path that minimizes the sum of distances between all pairs of elements, the technique can be extended to handle vector quantities by using euclidean distance between vectors. HMM are statistical technique that creates a parametric model based on observations from a process evolving in time, the likelihood of any other sequence of being generated by the model can then be calculated by using the trained model.

Most of the works based on the use of DTW to recognize quality of movement do not present any experiment to validate the performance of the method when applied to sequences that deviate slightly from the correct form. In this work we use experimental data collected among 14 subjects who were instructed to perform as set of physical activities according to the specifications of an expert,data from ten different subjects who were instructed to deviate from the correct performance for  the activities was also collected. This was used to test two evaluation approaches, one  based on Multi-Dimensional Dynamic Time Warping and another based on Hidden Markov Models. Results show that a system based on MDDTW is for the most part unable to correctly reject sequences of movement that deviate from the specification given by an expert in physical therapy when this deviation is small, and that a system based on HMM is far superior when it comes to evaluation of movement under these circumstances, provided that the system is trained with a set of sequences with low noise and that resulted from an accurate tracking.

The rest of the paper is organized as follows: in Section II proposals to evaluate the quality of human movement are reviewed, Section III describes the application of HMM and DTW to the collected data, in order to model the activities of interest,  it also describes the calculations performed to calculate the characteristics used as inputs to the models. Section IV describes the experiments performed to test the techniques. Section V presents the result of using the techniques to try to detect  errors in the activities. Section VI discusses the findings based on the performance results.

\section{Related Work}
\label{related}
This section reviews proposals aimed at evaluating the quality of movement in human beings.

Different approaches have been proposed to deal with the evaluation problem. Staab\parencite{staab2014recognizing} defines a limited set of errors for each of the movements considered, and trains Support Vector Machines to detect these errors using data coming from a Kinect sensor. This approach however is hardly scalable, due to the difficulty in obtaining data for all the possible mistakes that can arise while performing the movements.

Another approach is the use of finite state machines, which encode the different states that the user is supposed to pass in order for an activity to be considered correctly executed. This approach is studied in \parencite{velloso2013qualitative} as an alternative to machine learning algorithms. This is also proposed by \parencite{ravi2013automatic} by using quaternion data and  Spherical Linear Interpolation (SLERP). The sequence of quaternions is compared to the result of an interpolation and great deviations are considered signs that the movement does not satisfy the specifications. The author does not show results for movements of the lower limbs. The conditions for which a system based on finite state machines can be useful for recognition are studied in \parencite{maung2013games}.

Statistical techniques have also been proposed, the goal of such systems is to train a model with normal performances of the activities and detect deviations from normality in subsequent instances of the activities. Paiement et al. \parencite{paiement2009online} train a model based on parzen windows to detect deviations from normality in people moving on stairs. Experimental thresholds are determined to decide when the activity does not correspond to normality. The main advantage of these techniques lies in the fact that it is only necessary to provide data for normal execution of the activities of interest. In the context of rehabilitation however a system  based on this technique would be unable to determine the exact cause of error for the repetition of the activity.

Some other systems are based on the DTW algorithm, a mathematical technique to compute the degree of similarity between two time series.Examples of these include the works of Su et al\parencite{su2012ensuring}, the authors consider only movement sequences of the hands and use fuzzy logic to evaluate performance. Cuellar et al. \parencite{cuellar2015approach} develop a system based on DTW for which the score depends on the distance found between an observed performance of the activity and a model recorded by a physiotherapist. These systems are based on a distance metric calculated on sequences of varying length, and the goal is to determine if a new sequence is close enough to the sequence that can be seen as template for the activity. We propose the use of HMM to model time series arising from the performance of the activities, and test the ability of such models to detect and reject sequences of movements that deviate slightly from those considered correct. This method is compared to MDDTW, for the task of determining if a sequence is close enough to a standard so as to consider it a correct repetition.

\section{Methods }
This section shows the calculations performed in order to characterize the movements taken into account in the database collected, it also presents the algorithms used for evaluation of performance as well as the mechanism used to determine whether a sequence is considered to be correct or not.

\begin{figure}[H]
\centering
\includegraphics[scale=0.3]{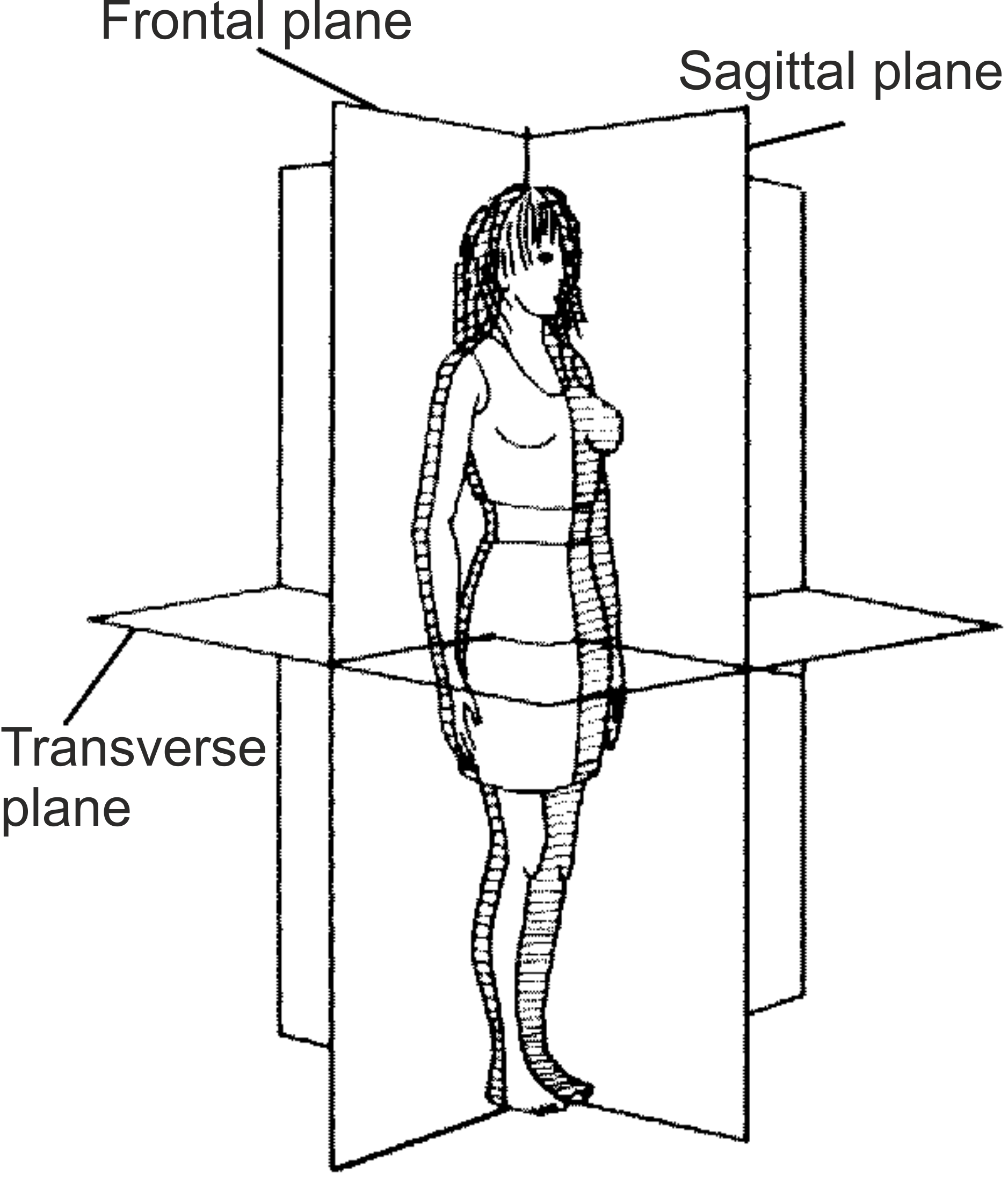}

\centering
\caption{Planes of movement}
\label{Planes}
\end{figure}

Movements in the fields of physical therapy and rehabilitation are defined in terms of three planes perpendicular to each other, known as the frontal, sagittal and transverse planes, as shown in Figure \ref{Planes}.

By using these planes, common movements are defined. Abduction is motion in a frontal plane and around an anteroposterior
axis that moves the segment away from the anatomical position.
Flexion of a joint results in segmental motion in a sagittal plane, around a mediolateral axis, and away from the anatomical position. Extension returns the segment to the anatomical position in a sagittal plane around a mediolateral axis and is described as increasing the angle at the joint. Internal rotation at the shoulder is motion of the arm segment around a superoinferior axis that rotates it from the palms-forward anatomical position to a posture in which the palms are facing more medially and finally posteriorly. External rotation is the opposite motion in which the segment is returned to or beyond the anatomical position in a transverse plane \parencite{mclester2007applied}.

\subsection{MDDTW}

Multi-Dimensional  Dynamic Time Warping is an extension of the DTW algorithm in which the warping path and the distance between two sequences are computed by taking the distance between two vectors on each step of the algorithm. This allows the use of a  pattern defined as a trajectory in  an hyperspace instead of a single valued function.

Given a set of trajectories considered as correct movements  $ {X_{1},X_{2}, ... , X_{N}} $, the vectors describing the points in such trajectories are normalized by subtracting the coordinates of the shoulder center and dividing by the distance between the shoulders for the case of upper limb movements, and subtracting the coordinates of the hip center and dividing by the distance between the hips for the case of lower limb movements. Once this is done MDDTW is calculated for all the possible pairs of sequences. 

Then a trajectory is selected for which the sum of these distances is minimal, this is  considered the template for the activity under consideration. 

Once this is done, the distance of all the sequences to the template is calculated, and an interval constructed around the mean of this distance. For evaluation the distance of the test sequences to the template is calculated, and those whose distance lies within the interval are considered to be close enough to the correct repetitions to be also considered correct.

\subsection{HMM}

Hidden Markov Models are stochastic models that consider an observed signal as the result of the transition of a system between several states, in each of these states there is a certain probability that one symbol might be observed. They have been  used in speech processing \parencite{rabiner1989tutorial}, and outlier detection\parencite{yan2013outlier} ,\parencite{zhu2015hmm}.

An HMM  model is totally specified when the transition matrix A, the observation  matrix B and the vector of initial states $\pi$ are known. 
We briefly discuss the structure of such matrices.

If a system has $N$ different possible states ${S_{1}, S_{2},...S_{N} }$ and in each state one of $M$ different symbols can be observed, the elements of the transition matrix $A$ describe the probability of passing from one state to another.

\begin{equation}
 a_{ij}=P[q_{t}=S_{j}|q_{t-1}=S_{i}]\quad  i\leq N \quad
j\leq N
\end{equation}

Where $q_{t}$ is the state of the system at time t.

The observation matrix $B$ describes the probability of observing a certain symbol while the system is in each of its states.

\begin{equation}
b_{ij}=P[v_{j} \quad  at  \quad  t|q_{t}=S_{i}]  \quad  i\leq N \quad
j\leq M
\end{equation}

where ${v_{1},v_{2},...,v_{M}}$ is the set of all the possible symbols that can be emitted by the system. 

Finally the initial state distribution $\pi = {\pi_{i}}$ gives the probabilities of the system being in one of the states for the initial observation.

\begin{equation}
\pi_{i}=P[q_{1}=S_{i}] \quad i\leq N
\end{equation}

In order to specify a certain movement by using HMMs we first determine three characteristics of the movement which are the angles formed with each of the three movement planes shown in Figure \ref{Planes}.To accomplish this, vectors normal to the planes must be estimated first by using the data coming from the sensor.

A vector normal to the transverse plane is given by the SDK.

\begin{equation}
V_{transverse}=(a,b,c,d)  .
\end{equation}

where

\begin{equation}
aX+bY+cZ+d=0;
\end{equation}

is the equation of the floor plane.

A vector normal to the frontal plane can be estimated by using the coordinates of the shoulders.
\begin{equation}
V_{frontal}=(p_{sc}-p_{ls})\times(p_{rs}-p_{ls})  .
\end{equation}

Where $p_{sc}$, $p_{ls}$ and $p_{rs} $ are the coordinates of the shoulder center, left shoulder and right shoulder as given by the sensor.

A vector normal to the sagittal plane is estimated using the previously calculated  vectors.

\begin{equation}
V_{sagittal}=V_{frontal}\times\ V_{transverse}  .
\end{equation}

The angles formed between the limb of interest and the planes of motion are calculated using these vectors as follows:

\begin{equation}
Frontal=90-\arccos{\frac{limb\cdot V_{frontal}}{\left \| limb \right \|\left \|  V_{frontal}\right \|}}  .
\end{equation}
\begin{equation}
Transverse=90-\arccos{\frac{limb\cdot V_{transverse}}{\left \| limb \right \|\left \|  V_{transverse}\right \|}}  .
\end{equation}
\begin{equation}
Sagittal=90-\arccos{\frac{limb\cdot V_{sagittal}}{\left \| limb \right \|\left \|  V_{sagittal}\right \|}}  .
\end{equation}

 Once the sequences of angles are obtained, they need to be quantized in order to use the discrete version of HMM, this quantization is done according to Table \ref{Quantization}. For any angular value there exists a symbol assigned to it, so that the sequence of angles becomes a discrete sequence, all of the possible values of the calculated angles are considered, so that any observed sequence can be assigned a certain probability. In order to account for the noise in the measurement a variation of ten degrees was considered as sufficient to change the emitted symbol.

\begin{table}[H]
\centering
\caption{Quantization of angles formed with the planes of motion}
\label{Quantization}
\begin{tabular}{l c l c l c l c l c l}
Range         & Symbol &    & Range       & Symbol \\
\hline
{[}-90 -80{]} & 1    & \hspace{1cm}  & {[}0 10{]}   & 10      \\
{[}-80 -70{]} & 2     & \hspace{1cm}  & {[}10 20{]}  & 11     \\
{[}-70 -60{]} & 3     & \hspace{1cm}  & {[}20 30{]}  & 12     \\
{[}-60 -50{]} & 4     & \hspace{1cm}   & {[}30 40{]}  & 13     \\
{[}-50 -40{]} & 5     & \hspace{1cm}  & {[}40 50{]}  & 14     \\
{[}-40 -30{]} & 6     & \hspace{1cm}   & {[}50 60{]}  & 15     \\
{[}-30 -20{]} & 7     & \hspace{1cm}    & {[}60 70{]}  & 16     \\
{[}-20 -10{]} & 8     & \hspace{1cm}  & {[}70 80{]}  & 17     \\
{[}-10 0{]}   & 9     &  \hspace{1cm} & {[}80 90{]}  & 18     \\
              &       &  \hspace{1cm}  &               &              &
\end{tabular}
\end{table}

Baum -Welch algorithm is used to train the model, which consists in estimating the matrices A and B.

With the trained model all of the trained sequences are assigned a probability by using the forward algorithm, and an interval built around the mean of such probabilities, by taking two standard deviations.

Evaluation of new sequences proceeds as follows: error sequences are quantized and then assigned a probability by using the forward algorithm. If this probability lies on the interval built with the correct repetitions, the sequence is considered to be correct.

\section{Experimental Setup}
Experiments were conducted to determine the effectiveness of the proposed algorithms to reject sequences that constitute deviations from what is considered a correct execution of the movements of interest.

Correct repetitions of the activities were recorded by asking 14 subjects to perform each one of the ten physical activities according to the specification three times.The activities considered are exercises commonly found in physical therapy and rehabilitation routines. These activities are:

\begin{enumerate}
\item Shoulder Extension
\item Shoulder Flexion
\item Shoulder Abduction
\item Hip Extension
\item Hip Flexion
\item Hip Abduction
\item Shoulder Internal Rotation
\item Shoulder External Rotation
\item Elbow Flexion
\item Elbow Extension.
\end{enumerate}

This means that 42 repetitions were recorded. Some of these were excluded from the analysis based on the noise present in the signals, in order to obtain a pattern as well defined as possible.

A database of mistakes during the execution of activities was collected by asking ten persons to perform the incorrect form of the activity ten times. This gives us a total of 100 incorrect repetitions for each activity.

The mistakes are explained as follows: 

for the abduction exercises, \textit{Error1} consists in deviating the limb of interest towards the front of the body while performing the activity, increasing the angle formed with the frontal plane.\textit{Error 2} consists in deviating the limb towards the back of the body, decreasing the angle formed with the frontal plane.

 For extension and flexion exercises \textit{Error 1} consists in deviating the limb away from the body , increasing the value of the angle formed with the sagittal plane. \textit{Error 2} consists in moving the limb towards the body, decreasing the angle formed with the sagittal plane. 

 For shoulder rotation exercises \textit{Error 1} consists in deviating the limb towards the floor, decreasing the value of the angle formed with the transverse plane.\textit{Error 2} consists in deviating the limb upwards, increasing the value of the angle formed with the transverse plane.

  \begin{figure}[H]
	  \centering
    \includegraphics[width=0.9\linewidth ]{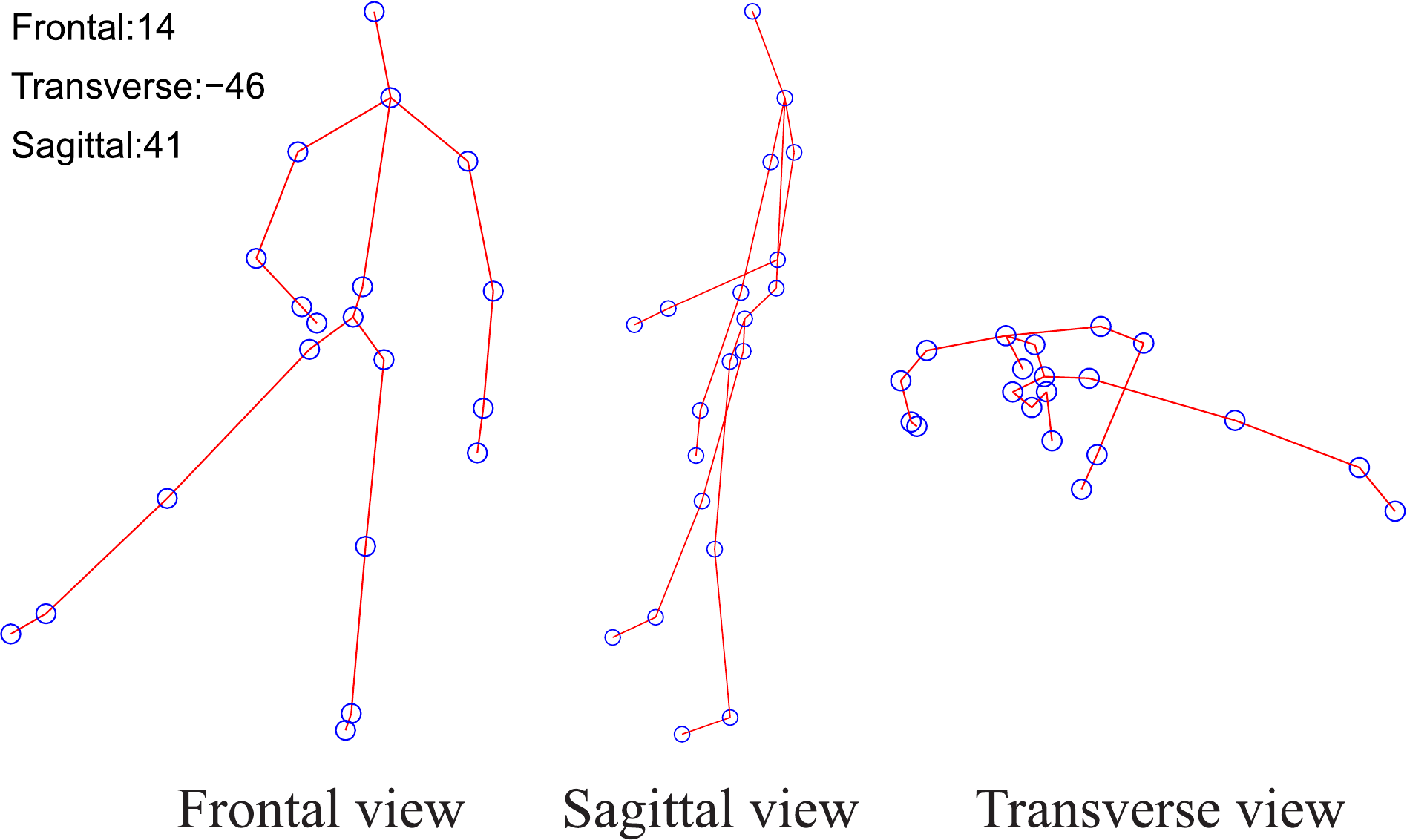}
    \caption{Erroneous performance consisting in moving the limb towards the front of the body}   
   \end{figure}
	
	\begin{figure}[H]
   \centering
	  \includegraphics[width=0.9\linewidth]{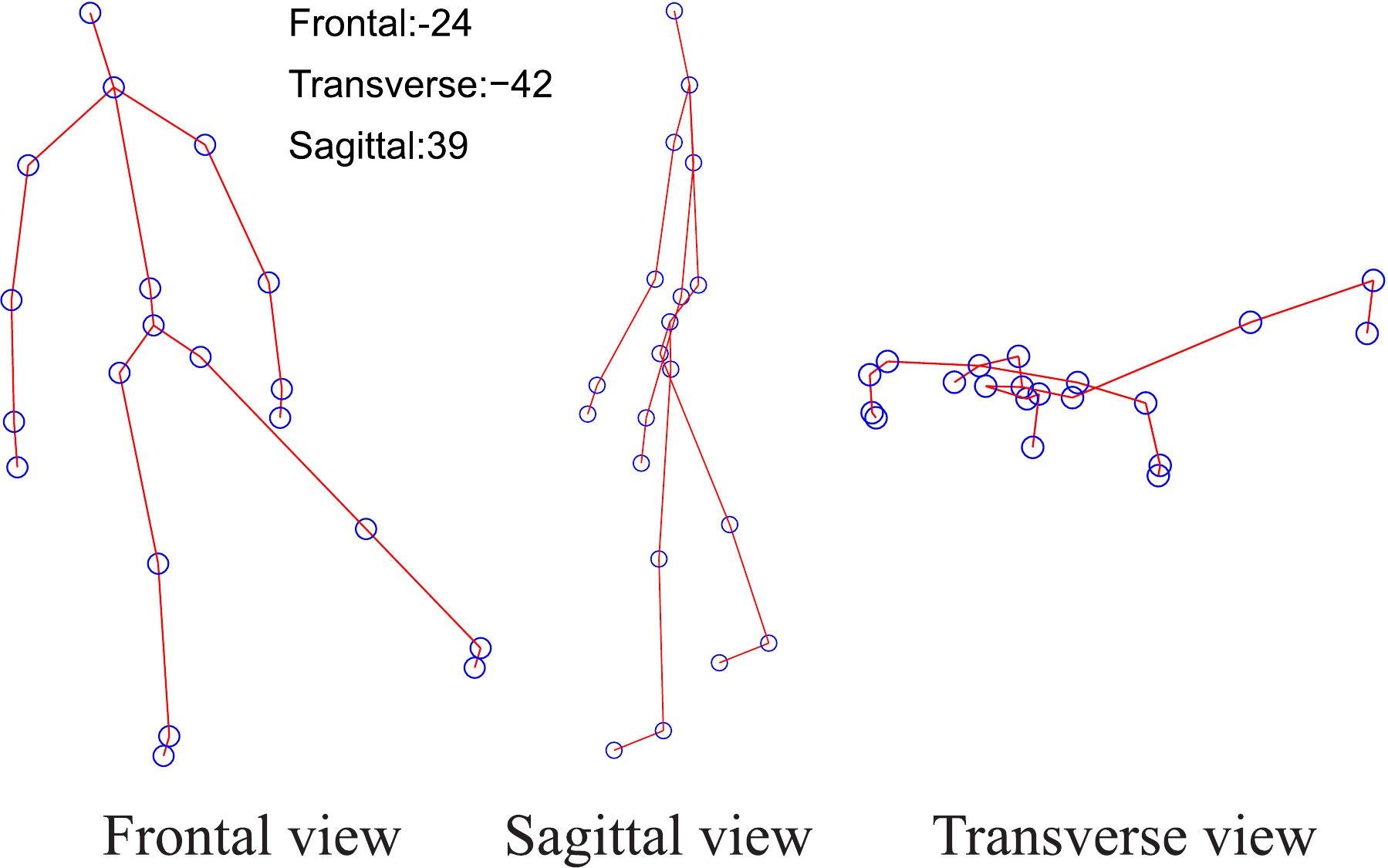}
	  \caption{Erroneous performance consisting in  moving the limb towards the back of the body}
	 \end{figure}

In our analysis of performance we have chosen to divide the activities performed in two basic movement phases, one that consists in the limb of interest moving away from the body and another that moves it back to the initial position. This allows evaluation to make sure that the movement adjusts closely to the specification.

The first experiment was done in order to test the ability of a DTW based system to reject movements that deviate from the specification. It consisted in determining using the correct repetitions for the activities both a template and an interval of distances to consider a repetition as correct. The data used in these experiments consists in coordinates of the limb of interest normalized to take into account relative displacement to the sensor and person size. The distance to the template is then calculated for all the incorrect repetitions on both phases of the movement and if it lies within the interval, the repetition can be considered correct.

The second experiment varied the data fed to the DTW recognition scheme, by using the estimated angles with respect to the planes of motion. A sequence of vectors containing the angles formed with these planes is used and a template calculated that minimizes the distance to the other sequences. Similar angular sequences are also calculated for the incorrect repetitions.

The third experiment consisted in training HMM models using the correct repetitions of the activities, the data used to train the model consists in the same coordinate sequences used for the first experiment. Since a model can be trained using discrete sequences, an HMM model is trained for each one of the characteristics of interest, for each one of the movement phases.This means that for each phase of an activity, three HMM models are generated.

The models were then tested with the incorrect repetitions, by using the forward algorithm, to determine the model's ability to reject them. 

The fourth experiment consisted in training HMM models varying the sequences used for training, this time the angular sequences  from experiment two are used. Once again an interval of probability values is obtained and the incorrect repetitions are tested by determining whether the probability assigned to them by the model lies within the interval.

\section{Results}

This section shows the recognition percentages for the errors taken into account for each one of the ten activities and for the two methods studied.

Table \ref{DTW} shows the results of the first experiment, the percentages correspond to erroneous sequences within the error database that were detected as such. Percentages of error recognition are shown for both the phases in which the movement has beed divided. Since the technique takes into account vectors used to describe the trajectories, only one metric distance is obtained for each error sequence and based on this distance the percentage of correctly refused erroneous sequences is calculated.

Table \ref{DTWangles} shows the results of the second experiment, percentages show the success of the technique in rejecting erroneous sequences executed by the participants.

Tables \ref{HMMangles1} and \ref{HMMangles2}  show the results of the third experiment. A different HMM model is trained for each one of the three angles of motion (sagittal, frontal and transversal). Three results are then shown for each one of the phases of movement and for each one of the types of errors considered. The highest result of the three can be considered the actual recognition of errors accuracy for HMM based models.

Tables \ref{HMMcoordinates1} and \ref{HMMcoordinates2} show the results  of the fourth experiment. Three models are trained, one for each one of the coordinates X,Y,Z of the sequence describing the activity. The highest percentage can be considered the recognition of deviation from normality for HMM models under these circumstances.

It is evident from tables \ref{DTW} and \ref{DTWangles} that the performance of MDDTW for recognizing deviations is reduced when using angular characteristics instead of normalized coordinates.

From Table \ref{Average} it can be seen that the performance of HMM is reduced when using coordinate sequences, in comparison to the use of  estimated angles formed with  the planes of motion.

It was found that the average error recognition rate for HMM  is consistently higher than for MDDTW, in both phases of the activities and for any characteristic, as can be seen in table \ref{Average}. This table shows the average success rate for each technique under each condition.

Generally speaking the best combination of characteristics and technique was using HMM models together with estimated angles formed with the planes of motion.

There were certain errors very hard to recognize with any combination of technique and characteristics, this is the case for an error 2 in the second phase of hip abduction (leg returning to its initial position), for which the maximum recognition percentage was merely 48,1\%.

\begin{table}[H]
\centering
\caption{Percentage of error detection using MDDTW with coordinate sequences}
\label{DTW}
\begin{tabular}{|l|l|l|l|l|}
\hline
                   & \multicolumn{2}{l}{Phase 1} & \multicolumn{2}{l}{Phase 2} \\ \hline
Activity           & Error 1      & Error 2      & Error 1      & Error 2      \\ \hline
Shoulder Abduction & 47           & 38,88        & 100          & 61,11        \\ \hline
Hip Abduction      & 18,75          & 48,10          & 15          & 29,11          \\ \hline
Shoulder Extension & 22,73        & 6            & 0            & 1            \\ \hline
Hip Extension      & 29,21        & 11           & 16,85        & 11           \\ \hline
Elbow Extension    & 47,37        & 50           & 46,32        & 50           \\ \hline
Shoulder Flexion   & 51           & 5,15         & 46           & 4,12         \\ \hline
Hip Flexion        & 20           & 12,12        & 24           & 13,13        \\ \hline
Elbow Flexion      & 47           & 71           & 44           & 60           \\ \hline
Internal Rotation  & 71,59            & 7,29           & 9,09         & 16,67        \\ \hline
External Rotation  & 52           & 40           & 49           & 14           \\ \hline

\end{tabular}
\end{table}

\begin{table}[H]
\centering
\caption{Percentage of error detection using MDDTW with angle sequences}
\label{DTWangles}
\begin{tabular}{|l|l|l|l|l|}
\hline
                   & \multicolumn{2}{l}{Phase 1} & \multicolumn{2}{l}{Phase 2} \\ \hline
Activity           & Error 1      & Error 2      & Error 1      & Error 2      \\ \hline
Shoulder Abduction & 34,29          & 7,78        & 62,9          & 18,89       \\ \hline
Hip Abduction      & 22,5          & 11,4          & 13,75          & 2,53          \\ \hline
Shoulder Extension & 1,13          & 1          & 0            & 1            \\ \hline
Hip Extension      & 3,37         & 0           & 4,5        & 2         \\ \hline
Elbow Extension    & 22,11       & 30,21          & 8,42        & 11,46           \\ \hline
Shoulder Flexion   & 12           & 1,03         & 3           & 1,03         \\ \hline
Hip Flexion        & 0           & 1,01        & 0           & 0        \\ \hline
Elbow Flexion      & 35          & 51           & 36           & 10           \\ \hline
Internal Rotation  & 0            & 0          & 1,14         & 0       \\ \hline
External Rotation  & 1           & 1          & 1          & 1           \\ \hline

\end{tabular}
\end{table}

\begin{table}[H]
\centering
\caption{Error recognition results for  HMM in the first phase of the movements using coordinate sequences}
\label{HMMcoordinates1}
\begin{tabular}{|l|l|l|l|l|l|l|}
\hline
    & \multicolumn{6}{l|}{\textbf{Phase 1}}       \\ \hline
\textbf{Activity}           & \multicolumn{3}{l|}{\textbf{Error1} \textbf{\%}} & \multicolumn{3}{l|}{\textbf{Error2} \textbf{\%} } \\ \hline

\textbf{Shoulder Abduction} & \cellcolor{blue!25}95.7    &  17.1 & 84.3 & 96.7 & 12.2 & \cellcolor{blue!25}98.9 \\ \hline
\textbf{Hip Abduction}      & 70  &  \cellcolor{blue!25}83.8  & 71.3  & 72.2 & \cellcolor{blue!25} 81,0  &  54,4  \\ \hline
\textbf{Shoulder Extension} &\cellcolor{red!25} 68.2 & 35.2 & 4.5& \cellcolor{red!25}56.0  & 37.0   & 16.0  \\ \hline
\textbf{Hip Extension}      & 73.03 & \cellcolor{blue!25}84.3 & 80.9 & 67.0  & \cellcolor{blue!25}83.0  & 80.0  \\ \hline
\textbf{Elbow Extension}   & \cellcolor{red!25} 36.8 & 7.3 & 14.7 & \cellcolor{red!25}62.5 & 13.5 & 14.6   \\ \hline
\textbf{Shoulder Flexion}   & \cellcolor{blue!25}89.0  & 23.0 & 75.0  & \cellcolor{blue!25} 81.4 & 18.6  & 69.1 \\ \hline
\textbf{Hip Flexion}        & \cellcolor{blue!25}81.0  & 0.0   & 1.0  & \cellcolor{red!25} 52.5 & 4.1 & 0 \\ \hline
\textbf{Elbow Flexion}      & \cellcolor{blue!25}94.0  & 66.0  &  49  & \cellcolor{blue!25}94.0 & 79.0  & 71.0  \\ \hline
\textbf{Internal Rotation}  & \cellcolor{red!25}50 & 46.6 & 1.1 & 37.5 & \cellcolor{red!25}51.0 & 2.1 \\ \hline
\textbf{External Rotation}  & 0  & \cellcolor{red!25}31.0  & 2.0  & 1.0   & \cellcolor{red!25}41.0  & 2.0 \\ \hline
\end{tabular}
\end{table}

\begin{table}[H]
\centering
\caption{Error recognition results for  HMM in the second phase of the movements using coordinate sequences}
\label{HMMcoordinates2}
\begin{tabular}{|l|l|l|l|l|l|l|}
\hline
    & \multicolumn{6}{l|}{\textbf{Phase 2}}       \\ \hline
\textbf{Activity}           & \multicolumn{3}{l|}{\textbf{Error1} \textbf{\%}} & \multicolumn{3}{l|}{\textbf{Error2} \textbf{\%} } \\ \hline

\textbf{Shoulder Abduction} &\cellcolor{blue!25} 94.3 & 61.43 & 75.7 & \cellcolor{blue!25}98.9 & 30 &  97.8 \\ \hline
\textbf{Hip Abduction}      & 50.0 &  40.0 & \cellcolor{red!25} 70.0 & 36,7 &  27.8 & \cellcolor{red!25}   48.1  \\ \hline
\textbf{Shoulder Extension} & \cellcolor{blue!25}67.0  & 50.0 & 1.1 & \cellcolor{red!25}56.0  & 44.0   & 4.0   \\ \hline
\textbf{Hip Extension}      & 69.7 & 74.2 &\cellcolor{red!25} 74,2 & 63.0  & \cellcolor{blue!25}77.0  & 70.0  \\ \hline
\textbf{Elbow Extension}    & \cellcolor{blue!25}14.7 & 7.4 & 10.5 & 15.6 & 12.5& \cellcolor{red!25}15.6 \\ \hline
\textbf{Shoulder Flexion}   & 74.0 & 5.0  &\cellcolor{blue!25} 88.0 & 73.2 & 4.1 & \cellcolor{blue!25}82.5 \\ \hline
\textbf{Hip Flexion}        & \cellcolor{red!25}52.0   & 0.0   & 2.0  & \cellcolor{red!25} 34.3  & 3.0 & 5.0 \\ \hline
\textbf{Elbow Flexion}      & \cellcolor{blue!25} 84.0   & 6.0  & 73.0  & \cellcolor{blue!25} 91.0 & 7.0  & 86.0      \\ \hline
\textbf{Internal Rotation}  & 28.41 & \cellcolor{red!25}36.4 & 1,1 & 26.0 & \cellcolor{red!25}35.4 & 2.1 \\ \hline
\textbf{External Rotation}  & 0.0  & \cellcolor{blue!25}75.0  & 3.0  & 2.0  & \cellcolor{blue!25}81.0 & 9.0      \\ \hline
\end{tabular}
\end{table}

\begin{table}[H]
\centering
\caption{Error recognition results for  HMM in the first phase of the movements using angle sequences}
\label{HMMangles1}
\begin{tabular}{|l|l|l|l|l|l|l|}
\hline
    & \multicolumn{6}{l|}{\textbf{Phase 1}}       \\ \hline
\textbf{Activity}           & \multicolumn{3}{l|}{\textbf{Error1} \textbf{\%}} & \multicolumn{3}{l|}{\textbf{Error2} \textbf{\%} } \\ \hline

\textbf{Shoulder Abduction} & 90.0    & \cellcolor{blue!25} 92.9 & 21.4 & 34.4 &\cellcolor{blue!25} 96.7 & 26.7 \\ \hline
\textbf{Hip Abduction}      &32.5  & \cellcolor{red!25} 62.5  & 41.3  & \cellcolor{red!25}36.7 &  30.4  &  8.9  \\ \hline
\textbf{Shoulder Extension} & 14.8 & 38.63 & \cellcolor{red!25}64.8 & \cellcolor{red!25}45.0  & 3.0   & 7.0  \\ \hline
\textbf{Hip Extension}      & \cellcolor{blue!25}78.7 & 40.4 & 71.9 & \cellcolor{red!25}72.0  & 30.0  & 25.0  \\ \hline
\textbf{Elbow Extension}    & 17.9 & 35.8 &\cellcolor{red!25} 54.7 & 26.0 & \cellcolor{red!25}57.3 & 21.9   \\ \hline
\textbf{Shoulder Flexion}   & 17.0  & 54.0  & \cellcolor{red!25}68.0  & 10.3 & 63.9  & \cellcolor{blue!25}77.3 \\ \hline
\textbf{Hip Flexion}        & 12.0  & 7.0   & \cellcolor{blue!25}99.0  & 27.3 & 16.2 & \cellcolor{red!25}29.3 \\ \hline
\textbf{Elbow Flexion}      & 89.0  & 89.0  & \cellcolor{blue!25}99.0  & \cellcolor{blue!25}90.0 & 81.0  & 25.0  \\ \hline
\textbf{Internal Rotation}  & 68.2 & \cellcolor{blue!25}77.3 & 32.9 & 43.3 & \cellcolor{blue!25}75.3 & 39.2 \\ \hline
\textbf{External Rotation}  & 16.0  & \cellcolor{blue!25}88.0  & 25.0  & 17.0   & \cellcolor{red!25}68.0  & 38.0 \\ \hline
\end{tabular}
\end{table}

\begin{table}[H]
\centering
\caption{Error recognition results for  HMM in the second phase of the movements using angle sequences}
\label{HMMangles2}
\begin{tabular}{|l|l|l|l|l|l|l|}
\hline
    & \multicolumn{6}{l|}{\textbf{Phase 2}}       \\ \hline
\textbf{Activity}           & \multicolumn{3}{l|}{\textbf{Error1} \textbf{\%}} & \multicolumn{3}{l|}{\textbf{Error2} \textbf{\%} } \\ \hline

\textbf{Shoulder Abduction} & 77.1 & 75.7 & \cellcolor{blue!25}82.9 & 52.2 & 78.9 & \cellcolor{blue!25} 92.2 \\ \hline
\textbf{Hip Abduction}      & 22.5 &  \cellcolor{red!25}60.0 &  41.2 & 10.1 &  \cellcolor{red!25}21.5 &   7.6  \\ \hline
\textbf{Shoulder Extension} & 25.0  & 36.4 & \cellcolor{blue!25}76.1 & \cellcolor{red!25}51.0  & 1.0   & 9.0   \\ \hline
\textbf{Hip Extension}      & 21.3 & 28.9 &\cellcolor{blue!25} 95.5 & 37.0  & 24.0  &\cellcolor{red!25} 57.0  \\ \hline
\textbf{Elbow Extension}    & 63.1 & 28.4 & \cellcolor{blue!25}87.4 & 79.2 & 32.3 & \cellcolor{blue!25}98.9 \\ \hline
\textbf{Shoulder Flexion}   & 33.0 &\cellcolor{red!25} 72.0  & 25.0 & 30.9 & \cellcolor{blue!25}78.3 & 47.4 \\ \hline
\textbf{Hip Flexion}        & 0   & 7.0   & \cellcolor{blue!25}97.0  & 22.2  & 11.1 & \cellcolor{red!25}28.3 \\ \hline
\textbf{Elbow Flexion}      & 19.0   & 46.0  & \cellcolor{blue!25}78.0  & 18.0 & \cellcolor{blue!25}78.0  & 55.0      \\ \hline
\textbf{Internal Rotation}  & 48.9 & \cellcolor{blue!25}73.9 & 27.3 & 39.2 & \cellcolor{blue!25}77.3 & 24.7 \\ \hline
\textbf{External Rotation}  & 30.0  & \cellcolor{blue!25}91.0  & 43.0  & 38.0  & 38.0 & \cellcolor{red!25}59.0      \\ \hline
\end{tabular}
\end{table}

\begin{table}[H]
\centering
\caption{Average recognition percentages for each technique for each combination of factors}
\label{Average}
\begin{tabular}{|l|l|l|l|l|}
\hline
\multirow{8}{*}{Phase 1} & \multirow{4}{*}{Coordinates} & \multirow{2}{*}{Error 1} & HMM & 71,3 \\ \cline{4-5} 
                         &                              &                          & DTW & 40,6 \\ \cline{3-5} 
                         &                              & \multirow{2}{*}{Error 2} & HMM & 70,1 \\ \cline{4-5} 
                         &                              &                          & DTW & 28,9 \\ \cline{2-5} 
                         & \multirow{4}{*}{Angles}      & \multirow{2}{*}{Error 1} & HMM & 78,1 \\ \cline{4-5} 
                         &                              &                          & DTW & 13,1 \\ \cline{3-5} 
                         &                              & \multirow{2}{*}{Error 2} & HMM & 64,7 \\ \cline{4-5} 
                         &                              &                          & DTW & 10,4 \\ \hline
\multirow{8}{*}{Phase 2} & \multirow{4}{*}{Coordinates} & \multirow{2}{*}{Error 1} & HMM & 65,5 \\ \cline{4-5} 
                         &                              &                          & DTW & 35,0 \\ \cline{3-5} 
                         &                              & \multirow{2}{*}{Error 2} & HMM & 62,0 \\ \cline{4-5} 
                         &                              &                          & DTW & 26,0 \\ \cline{2-5} 
                         & \multirow{4}{*}{Angles}      & \multirow{2}{*}{Error 1} & HMM & 81,3 \\ \cline{4-5} 
                         &                              &                          & DTW & 13,0 \\ \cline{3-5} 
                         &                              & \multirow{2}{*}{Error 2} & HMM & 64,7 \\ \cline{4-5} 
                         &                              &                          & DTW & 47,9 \\ \hline
\end{tabular}
\end{table}

\section{Conclusion}

Experiments were conducted to test the ability of two algorithms to detect deviations from normality during the execution of activities common in physical therapy routines. The movements tested comprise activities for both the upper and lower limbs, and were Shoulder Abduction, Hip Abduction, Shoulder Extension, Hip Extension, Elbow Extension, Shoulder Flexion, Hip Flexion, Elbow Flexion, Internal Rotation and External Rotation. For each one of the movements sequences corresponding to correct executions and deviations from these were recorded. The deviations considered afected the position of the limb only with respect to one of the planes of motion. For abduction movements the deviation was introduced with respect to the angle formed with the frontal plane, for flexion and extension movements the deviation was introduced with respect to the angle formed with the sagittal plane, and for rotation movement the deviation was introduced with respect to the transverse plane.

While MDDTW is a useful tool to determine the similarity between time series coming from observations of human beings executing movements, has proven unable to reject sequences of movement which are similar to the standard but deviate in one of the directions relative to the planes of motion.

HMM has proved to be superior to  MDDTW for the task of detecting deviations in  movement sequences defined for physical therapy and rehabilitation. This is due to the ability of these models to penalize heavily those sequences of symbols that contain symbols not present in the sequences that were used to train the model, and those sequences containing symbols that imply a very unlikely transition between states.

For shoulder abduction, rotation and flexion results of the combination of HMM and angles with the planes of motion were satisfactory. For shoulder and elbow extension and hip abduction the proposed technique does not perform well, probably due to the noise  added with loss of precision of the sensor with distance.

It remains interesting as future work to determine if the use of quaternion data coming from the sensor might yield even better results for the task of detecting deviations from normality. Another future work might be the use of a discriminative model such as Conditional Random Fields to be used as outlier detectors.

\printbibliography

%\begin{acknowledgements}
%\end{acknowledgements}


\end{document}